\documentstyle[prd,aps,preprint,epsf,floats]{revtex}
\draft
\tighten

\def \MSbarbasic {\overline{{\rm MS}}}
\def \MSbar {\ifmmode \MSbarbasic \else $\MSbarbasic$\fi }
\newcommand{\cfig}[2]{\hbox to \columnwidth {\hfil\epsfysize = #2 \epsfbox{#1}
\hfil}}

\begin{document}

\preprint{
        \parbox{1.5in}{%
           hep-ph/9704344 \\
           PSU/TH/182 \\
           CERN-TH/97-73
        }
}

\title{Direct Estimation of Sizes of Higher-Order Graphs}

\author{John C. Collins$^{a,b}$,
        Andreas Freund$^{a}$
}

\address{
        $^{a}$Department of Physics, Penn State University,
            104 Davey Lab., University Park, PA 16802, U.S.A.
\\
        $^{b}$CERN --- TH division, CH-1211 Geneva 23, Switzerland.
        \thanks{Until 30 April 97.}
}

\maketitle

\begin{abstract}
    With the aid of simple examples
    we show how to make simple
    estimates of the sizes of higher-order Feynman graphs.
    Our methods enable appropriate values of renormalization and
    factorization scales to be made.
    They allow the diagnosis of the source of unusually large
    corrections that are in need of resummation.
\end{abstract}

\vfill
\noindent
CERN-TH/97-73 \\
17 April 1997

\section{Introduction}
\label{sec:introduction}

The starting point for this paper is formed by the following
observations:
\begin{itemize}

\item
    The only (known) systematic method for calculating scattering
    in QCD is perturbation theory.  (Lattice Monte-Carlo methods
    work in Euclidean space-time, and are excellent for
    calculating static quantities such as masses from first
    principles.  But they are essentially useless when a
    calculation in real Minkowski space-time is needed.)

\item
    In field theory, calculations beyond low orders of
    perturbation theory are computationally complex, both because
    the calculations of individual graphs are hard and because
    there are many different graphs.

\item
    Hence it is important to make the most efficient use of
    low-order calculations.

\end{itemize}
Since the coupling in practical calculations is not very weak,
the accuracy of predictions can be ruined by uncalculated
higher-order terms.  It follows that there is a need to estimate the
sizes of the errors.
For this one wants quick estimates of terms in perturbation
theory.  The computational complexity of the estimates should
increase as little as possible with the size of the graphs.
Indeed, our aim is that one only calculates integrals of the form
\begin{equation}
      \int _{l}^{u} dx \, x^{n} \, \ln^{p}x .
\end{equation}
With suitable methods:
\begin{itemize}

\item
    One can determine good values for renormalization and
    factorization scales, by asking how to minimize the error
    estimates.

\item
    When the estimates get substantially larger than some
    appropriate ``natural'' size, one would get a diagnosis of a
    need for resummation of classes of higher-order corrections.
    The diagnosis would include an explanation of the large terms
    and thus indicate the physics associated with the
    resummation.

\end{itemize}

In this paper, we explain how to start such a program.  It
builds on work first reported in Ref.\ \cite{Durham}.  Our
methods treat properties of the integrands of Feynman graphs, and
are therefore directly sensitive to the physics of the process
being discussed.  Some other treatments of these issues discuss
the problems in terms of the mathematics of series expansions in
general, without asking what is causing the graphs to be the
sizes they are.  A particular exception is the work of Brodsky,
Lepage, and Mackenzie \cite{BLM}\footnote{
   See also the more recent work of Brodsky and Lu \cite{BL} and
   of Neubert \cite{Neubert}.
}.
They use heavy quark loops to
probe the actual momentum scales that dominate in a particular
calculation; this is then used to provide a suitable value for
the renormalization/factorization scale.
But we believe that our methods provide a more direct route to
answering the question of why the scales are what they are and
why a calculation gives a particular order of magnitude.
The issues addressed by methods involving the Borel transform and
Pad\'e summation address complementary issues \cite{Pade}.

Now, most cross sections in QCD cannot be directly computed by
perturbation theory; this can only be used to
compute the short-distance coefficients that appear in the
factorization theorem.  So we will also treat the specific problems
that arise in estimating the sizes of the coefficient functions.
These functions have the form of a sum over Feynman graphs
(typically massless), with subtractions to cancel some infra-red
(IR) divergences. Remaining IR divergences are cancelled between
different graphs or between different final-state cuts. Two
characteristic features appear.  First, we can obtain estimates
for sums of particular sets of graphs, but not for the individual
graphs, which are divergent. Secondly, the coefficient functions
are not in fact genuine functions. They are normally singular
generalized functions (or distributions) and an estimate can only
be made for the integral of a coefficient function with a smooth
test function.

We show how to estimate the sizes of graphs by a direct
examination of the integrands.
An important part of our technique is an implementation of
subtractions directly in the integrands,
both for the subtractions that implement counterterms for
ultra-violet (UV)
renormalization and for the IR subtractions that are used in
short-distance coefficient functions.
In order to explain our ideas, we will examine two examples:
(1) the one-loop self-energy graph in $(\phi ^{3})_{4}$ theory,
and (2) a particular set of graphs for the Wilson coefficient for
deep-inelastic scattering.
Our estimates are in the form of approximations to
ordinary integrals that are absolutely convergent.
This is in contrast to the original integrals, which
are typically divergent in the absence of a regulator.
Thus a by-product of our work will be algorithms
for computing graphs numerically in Minkowski space-time, which
may have relevance to work such as Ref.\ \onlinecite{Catani.Seymour}.
As an illustration of how estimations can be carried out, even
analytically, for a measurable quantity, we will estimate the size of
the Wilson coefficient for the structure functions $F_{T}$ and $F_{L}$
in the last section.

\section{Euclidean self-energy in $\phi ^{3}$ theory}
\label{sec:self.energy}

\begin{figure}
\cfig{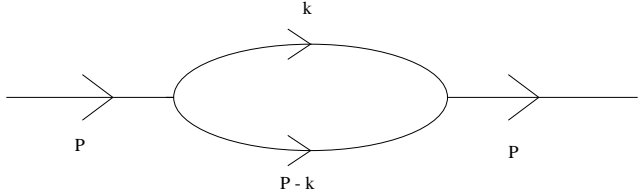}{1in}

\caption{One-loop self-energy graph.}
\label{fig:self.energy}
\end{figure}

In this section, we give a representation of the one-loop
self-energy graph of
Fig.\ \ref{fig:self.energy} in $\phi ^{3}$ theory in
four dimensions.
A particular renormalization scheme is used, which we relate to
ordinary \MSbar{} renormalization.
Then we show how to estimate the size of the graph from
elementary integrals and hence how to choose the renormalization
scale suitably.

\subsection{Renormalization}

The only complication in computing the graph of Fig.\
\ref{fig:self.energy} is its UV
divergence, given that we
choose to work in Euclidean space.
We give a
representation of the graph that is an absolutely convergent
integral over the loop momentum itself in four dimensions and in
which the renormalization is explicitly ``minimal''.  This last
term means that the counterterm for the (logarithmic) divergence is
independent of the mass and the external momentum.  It is a
very useful property when one wants to take zero-mass limits,
etc.  For a general divergence, the counterterm would be
polynomial in masses and external momenta.

Our representation of the graph is:
\begin{equation}
   I(p) = \frac {1}{2} \frac {g^{2}}{(2\pi )^{4}} \int  d^{4}k
          \left\{
            \frac {1}{(k^{2}+m^{2}) \, \left[ (p-k)^{2}+m^{2}\right]}
            - \frac {\theta (k>\mu _{c})}{k^{4}}
          \right\} .
\label{Ren.self.energy}
\end{equation}
We recognize in the integrand a term that is given by the usual
Feynman rules (in Euclidean space-time), and a subtraction term.
The subtraction term is the negative of the asymptote of the
first term as $k\to \infty $, so we term our procedure\footnote{
    See \cite{Durham} for a previous account.  A formalization of
    such ideas (to all orders of perturbation theory) was given
    earlier by Ilyin, Imashev and Slavnov \cite{Slavnov}, and
    later by Kuznetsov and Tkachov \cite{KT}.
}
``renormalization
by subtraction of the asymptote''.  A cut-off is applied to
prevent the subtraction term giving an IR divergence at
$k=0$; the cut-off does not affect the $k\to \infty $ behavior and
therefore does not affect the fact that the UV divergence is
cancelled.
To see that Eq.\ (\ref{Ren.self.energy}) is
equivalent to standard renormalization, one simply applies a
UV regulator, after which each term can be integrated
separately.  The first term is the unrenormalized graph and the
second term is a $p$-independent counterterm.

Evidently, the integral is absolutely convergent, and
can therefore be computed by any appropriate numerical method.
(It can also be evaluated analytically.  But this is not
interesting to us, since we wish to obtain methods that work for
integrals that are too complicated for purely analytic methods to
be convenient or useful.)

The counterterm is in fact the most general one that is
independent of $m$ and $p$, since any other renormalization
counterterm can differ only by a finite term added to the
integral that is independent of $m$ and $p$, and a change of the
cut-off $\mu _{c}$ is equivalent to adding such a term.  We can relate
the counterterm to the commonly used \MSbar{} one simply
by computing the counterterm alone, with dimensional
regularization:
\begin{equation}
    \mbox{standard pre-factor} \times
    \int _{|k|>\mu _{c}} d^{n}k \, \frac {1}{k^{4}} .
\end{equation}
The result is that setting $\mu _{c}$ equal to the scale $\mu $ of the
\MSbar{} scheme gives exactly \MSbar{} renormalization.  In
general, we would find that $\mu _{c}$ would be a factor times $\mu $, or
equivalently that we should set $\mu _{c}=\mu $ and then add a specific
finite counterterm to the graph.

Notice that the integral to relate our renormalization scheme to
the \MSbar{} scheme is algorithmically simpler to compute
analytically than the original integral.
There is always the possibility of adding finite counterterms.
Moreover, the precise form of the cut-off is irrelevant to the
general principles.  One can, for example, change the sharp
cut-off function $\theta (k>\mu _{c})$ to a smooth function $f(k/\mu _{c})$
that
obeys $f(\infty )=1$ and $f(0)=0$.  Such a function would probably be
better in numerical integration.

Of course, our method as stated is specific to one-loop graphs.
But it is an idea that has been generalized \cite{Slavnov,KT} to
higher orders.

\subsection{Estimate}

We now show how to estimate the size of the integral Eq.\
(\ref{Ren.self.energy}).  To give ourselves a definite case, let
us choose $p^{2} \lesssim m^{2}$.  We obtain the estimate as the sum of
contributions from $k<m$ and from $k>m$.  Since the
renormalization counterterm is designed to subtract the $k\to \infty $
behavior of the unrenormalized integrand, we regard it as a
$\delta $-function at infinity and therefore to be associated
completely with the $k>m$ term in our estimate.

In the region $k<m$, our estimate is obtained by replacing each
propagator by $1/m^{2}$ so that
\begin{eqnarray}
  \mbox{Contribution from $k<m$}
  &\simeq& \frac {g^{2}}{32\pi ^{4}} \int _{k<m} d^{4}k \frac {1}{m^{4}}
\nonumber\\
  &=& \frac {g^{2}}{64\pi ^{2}} .
\end{eqnarray}
This factor is the product of $g^{2}/32\pi ^{4}$ for the prefactor and
$\pi ^{2}/2$ for the volume of a unit 4-sphere.  As advertised, we have
had to calculate no integral that is more complicated than a simple
power of $k$.  The approximation of replacing the propagators by
$1/m^{2}$ leads us to an over-estimate of the integral, but not by a
great factor, since we are in a region of small momentum.

The estimate for $k>m$ is obtained by replacing the propagators
by their large $k$ asymptote:
\begin{eqnarray}
  \mbox{Contribution from $k>m$}
  &\simeq& \frac {g^{2}}{32\pi ^{4}} \int _{k>m} d^{4}k
            \left[ \frac {1}{k^{4}} - \frac {\theta (k>\mu _{c})}{k^{4}}
\right]
\nonumber\\
  &=& \frac {g^{2}}{32\pi ^{4}} 2\pi ^{2}
  \left[
     \int _{m}^{\infty }\frac {dk}{k} - \int _{\mu }^{\infty }\frac {dk}{k}
  \right] .
\end{eqnarray}
Since we are taking the difference of two terms, we must be
careful about the errors, which are of order
\begin{equation}
   \int _{k>m} d^{4}k \frac {m^{2}}{k^{6}} = \frac {g^{2}}{32\pi ^{2}}.
\end{equation}

To understand the structure of the result, let us examine how the
original integral (\ref{Ren.self.energy}) appears after
integrating over the angle of $k$:
\begin{equation}
    \frac {g^{2}}{16\pi ^{2}}
     \int _{0}^{\infty }\frac {dk}{k} \left[ A(k,p,m) - \theta (k>\mu _{c})
\right] .
\label{log.integral}
\end{equation}
The function $A$ is the angular average of $k^{4}$ times the two
propagators.  It approaches 0 as $k\to 0$, so that the integral
is convergent there, and it approaches unity as $k\to \infty $, which
would give a UV divergence were it not for the
subtraction.

\begin{figure}
\cfig{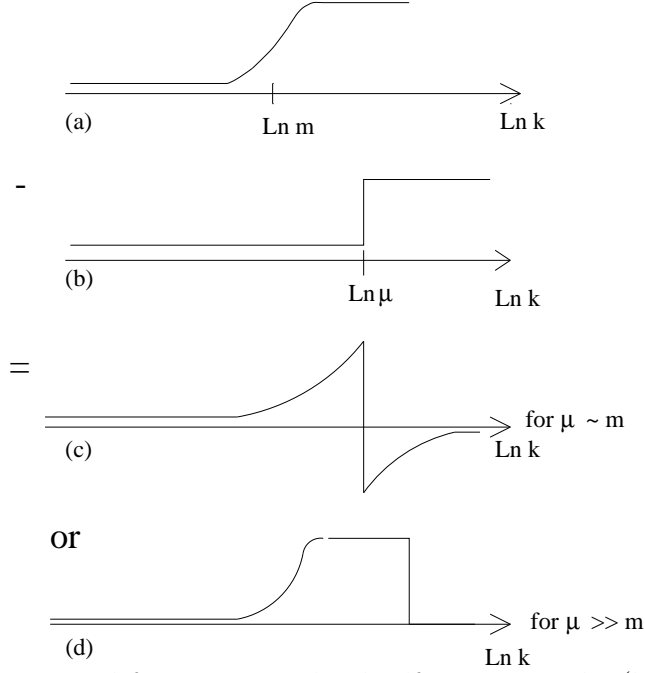}{3.5in}

\caption{(a) Scaled integrand for unrenormalized self-energy graph.
    (b) Subtraction term.
    (c) Total if $\mu \sim m$.
    (d) Total if $\mu  \gg m$.}
\label{fig:difference}
\end{figure}

We can represent this situation by the graphs of Fig.\
\ref{fig:difference}.  The transition region for the
unsubtracted integrand, where it changes from being 0 to 1,
is around $k=m$, to within a factor of 2 or so; this is evident
by examining the denominators.  By setting $\mu _{c}$ to be $m$ to
within a factor of 2, we achieve the following:
\begin{itemize}

\item
    The integrand is less than unity everywhere.

\item
    It is concentrated in a shell of thickness of order $m$
    around $k=m$.

\end{itemize}
Thus we can say that the natural size of the graph is the product
of
\begin{itemize}

\item The prefactor $g^{2}/32\pi ^{4}$.

\item The surface area of a unit 4-sphere: $2\pi ^{2}$.

\item The width of the important region, about unity, in units of
    $m$.

\item A factor of $m$ to the dimension of the integral, i.e., 1.

\item A factor less than unity, say $1/2$, to allow for the fact
    that, after subtraction, the integrand in Eq.\
    (\ref{log.integral}) is smaller than unity and that there is
    a cancellation between negative and positive pieces.

\end{itemize}
That is, the natural size is $g^{2}/32\pi ^{2}$, if $\mu $ is reasonably
close to $m$.

If $\mu $ is not close to $m$, then we get a long plateau in the
integrand---see Fig.\ \ref{fig:difference}.  The height of the
plateau is unity, and its length is $\ln (\mu /m)$ to about $\pm 1$ in
units of $\ln k$.  This clearly gives a larger-than-necessary
size, and an optimal choice of the \MSbar{} scale is around $m$.

One should not expect to get an exact value for the scale $\mu $.
A physical quantity in the exact theory is independent of $\mu $,
and any finite-order calculation differs from the correct value
by an amount whose precise value is necessarily unknown until one has done a
more accurate calculation.
If one is able to estimate the size of the error, as we are
proposing, then an appropriate value of $\mu $ is one that minimizes
the error.
Given the intrinsic imprecision of an error estimate, there is a
corresponding imprecision in the determination of $\mu $.
One can expect to identify, without much work,  an appropriate
scale $\mu $ to within a factor 2, and, with a bit more work, to
within perhaps $50\%$.  These estimates just come from asking
where the transition region in the integrand is, and by then
obtaining an answer by simple examination of the integrand.  But
one cannot enter into a religious argument of the wrong kind as
to whether the correct scale is $1.23m$ as opposed to $1.24m$,
for example.\footnote{
    Brodsky and Lu \cite{BL} obtain very precise estimates of a
    suitable scale.  Their rationale is the elimination of
    IR renormalons in the relations between IR-safe
    observables.  This is a concern with very high-order
    perturbation theory, an issue that we do not address.
}
By definition an error estimate is approximate.

Since we have not yet investigated how to estimate even
higher-order graphs, we are making the reasonable conjecture that the
properties of graphs do not change rapidly with order.  Then our
estimate that $\mu $ should be close to $m$ will ensure that
higher-order graphs are of the order of their natural size.

\subsection{Implication for QCD}

The same arguments applied to similar graphs in QCD show that the
natural expansion parameter in QCD is
\begin{equation}
   \frac {\alpha _{s}}{4\pi } \times  (\mbox{group theory})
      \times  (\mbox{factor for multiplicity of graphs}).
\label{natural.size.QCD}
\end{equation}
These arguments rely on being able to show that in the dominant part
of the range of integration all lines have approximately a
particular virtuality and that the relevant range of integration
is a corresponding volume of momentum space.

In the general case we cannot expect to get a much smaller
result, but we can expect that in some situations the properties
of the dominant integration region(s) will not be so good.
So what we need to do next is to analyze more interesting graphs
in QCD.  This we will do in the next section.

In general, when we get corrections in QCD that are substantially
larger than the natural size given above, it must be either
because the integrand is excessively large, or because there
is no single natural scale, or because we have not chosen a
good scale.\footnote{
    Our use of the word ``natural'' may suggest that we are
    proposing to estimate higher-order corrections simply by
    multiplying the appropriate power of the natural expansion
    parameter by the number of graphs.  This is not what we mean.
    We are arguing first that the sizes of graphs can actually be
    estimated fairly simply, and that the natural expansion
    parameter is a useful {\em unit} for these estimates.
    Secondly, we show that in the most favorable cases, graphs
    are less than or about unity in these natural units. Finally,
    we argue, in the next section, that general kinematic
    arguments about the physics of a graph are useful in
    diagnosing cases where graphs are large in natural units.
}
If the integrand is especially large relative to the natural
unit, or if there is no single natural scale, then we should
investigate in more detail the reasons, and derive something like
a resummation of higher-order corrections \cite{resum}.
In precisely such
situations, one does indeed have to compute high-order graphs. At
the same time, there is no need to compute the complete graphs
in all their gory detail, but only their simple parts.

\section{Wilson coefficient for deep-inelastic scattering}
\label{sec:Wilson}

The factorization formula for the leading-twist part of
a deep-inelastic structure function $F(x)$ is
\begin{equation}
   F(x) = \int  \frac {d\xi }{\xi } f(\xi ) \hat F(x/\xi ).
\end{equation}
Here, $f(\xi )$ is a parton density, and $\hat F$ is the
short-distance coefficient (``Wilson coefficient'').  We have
suppressed the indices for the different structure functions
($F_{1}$, $F_{2}$, etc.) and for the parton flavor.  The coefficient
function $\hat F(x/\xi )$ is obtained from Feynman graphs for
scattering on a parton target with momentum $\xi p$, where $p$ is
the momentum of the hadron target.  Subtractions
for initial-state collinear singularities
are applied to
the Wilson coefficient and the massless limit is taken.

\begin{figure}
\cfig{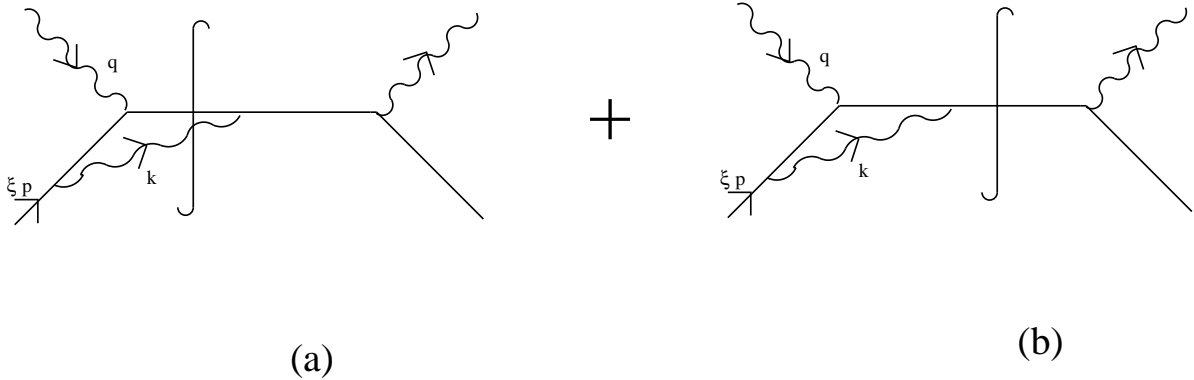}{2in}

\caption{The cuts of a one-loop graph for the Wilson coefficient
         for deep-inelastic scattering.}
\label{fig:1-loop.DIS}
\end{figure}

As an example, we will examine the contribution to the Wilson
coefficient from the  diagrams in Fig.\ \ref{fig:1-loop.DIS}.
These diagrams are the two possible cuts of a particular uncut
one-loop graph, and since we will need to use a cancellation of
final-state interactions, we must consider the sum of the two cut
graphs as a single unit.  Note:
\begin{itemize}

\item We will apply a subtraction to cancel the effect of
    initial-state collinear interactions where the incoming quark
    splits into a quark--gluon pair which are moving almost
    parallel to the incoming particle.\footnote{
     According to the factorization theorem, the subtractions cancel
     all the sensitivity to small
     momenta, i.e., to the initial-state collinear interactions.
     The subtractions are of the form of terms in the perturbative expansion
     of the distribution of a parton in a parton
     convoluted with lower-order terms in the coefficient
     function.  (See, for example, \cite{Colsop,Collins89} for details.)
     Of course, both the partonic cross section and the subtraction term
     have to be properly renormalized. We will call the subtraction terms
     eikonal because of the particular rules involved in their
     calculation, for graphs such as Fig.\ \ref{fig:1-loop.DIS}.
}

\item There will be soft-gluon interactions and collinear
    final-state interactions.  These will cancel after the sum
    over cuts.  (In a more general situation, a sum over a
    gauge-invariant set of graphs is necessary to
    get rid of all soft-gluon interactions.)

\item The Wilson coefficient is a distribution (or generalized
    function) rather than an ordinary function of $x/\xi $.  Thus it
    is useful to discuss only the size of the coefficient
    after it is integrated with a test function, but not the size
    of the unintegrated coefficient function.  The parton density
    $f(\xi )$ provides a ready-made test function that has a physical
    interpretation.

\item The first graph of Fig.\ \ref{fig:1-loop.DIS},
    which has a virtual gluon, has a 4-dimensional
    integral, but the second graph, with a real
    gluon, has only a 3-dimensional integral.  Thus the
    cancellations associated with the sum over cuts can only be
    seen after doing at least a 1-dimensional integral.

\end{itemize}

\begin{figure}
\cfig{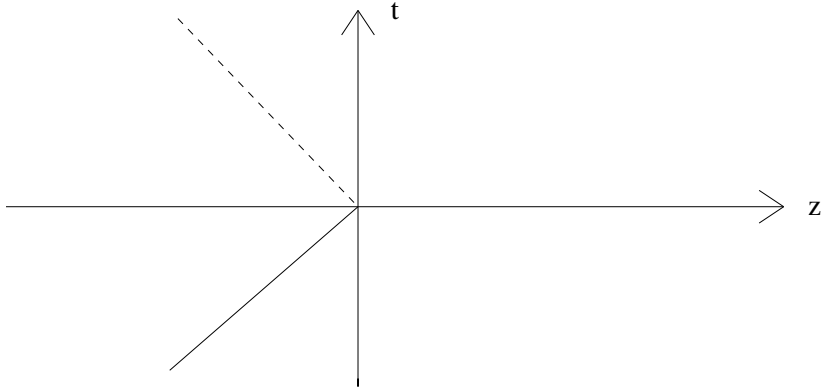}{3in}

\caption{Space-time structure of deep-inelastic scattering, in
    the center-of-mass frame of the virtual photon and the struck
    quark.  The solid line is the almost (light-like) world line
    of the incoming quark.  The dashed line is the world line of
    the single struck quark in the lowest-order (Born) graph for
    the hard scattering.
    }
\label{fig:sp-time}
\end{figure}

It is useful to visualize the process in space-time, Fig.\
\ref{fig:sp-time}, and to use light-front coordinates $(+,-,T)$
(defined by $V^{\pm }=(V^{0}\pm V^{3})/\sqrt 2$).  Our axes are such that the
incoming momenta for the hard scattering are:
\begin{equation}
  \xi  p^{\mu } = (\xi p^{+},0,{\bf 0}_{T}),
  ~~~~
    q^{\mu } = \left( -x p^{+},\frac {Q^{2}}{2xp^{+}},{\bf 0}_{T} \right) .
\end{equation}
Also, we find it convenient to parameterize the gluon momentum in Fig.\
\ref{fig:1-loop.DIS} in terms of two longitudinal momentum
fractions, $u$ and $z$, and a transverse momentum ${\bf k}_{T}$, as
follows:
\begin{equation}
    k^{\mu } = (u \xi (1-z) p^{+}, \, zq^{-}, \, {\bf k}_{T}).
\label{u.z.def}
\end{equation}
Thus $z$ is exactly the fraction of the total incoming minus
component of momentum that is carried off by the gluon, while $u$
is a scaled fraction of the plus component.
The scaling is somewhat unobvious, but it has the effect that positive
energy constraints on the final state restrict each of $u$ and
$z$ to range over the 0 to 1.

Now we summarize how the
calculation of the real and virtual graphs and of the
subtraction graphs \cite{Colsop}, contributing to the
Wilson coefficients, is to be carried out:
\begin{itemize}

\item Using the Feynman rules for cut diagrams, we write down the
     momentum integral with the appropriate
     $\delta $-functions.   Then we contract the trace over Dirac matrices
    with the appropriate transverse, longitudinal
    or asymmetric tensor on the photon indices to obtain the
    structure function $F_{1}$, $F_{2}$, etc.
    For the sake of a definite simple example
    we choose to contract with $-g_{\mu \nu }$,
    which in fact gives the combination $3F_{1} - F_{2}/2x$.

\item We express the results of the calculations in terms of
    the light-cone components of $k$ (the internal gluon
    momentum), of $p$ (the incoming
    quark momentum), and of $q$ (the incoming photon momentum).

\item In the graphs with real gluon emission, we use the two
    $\delta $-functions to perform the integrals over ${\bf k}_{T}$ and over
    the
    fractional momentum $\xi $ entering from the parton density.
    This leaves a 2-dimensional integral.

\item In the virtual graphs, we use the one $\delta $-function to
    perform the integral fractional momentum $\xi $ entering from
    the parton density.  We then perform the ${\bf k}_{T}$ integral
    analytically.  Again we have a 2-dimensional integral.

\item We change the integration variables to the scaled dimensionless
    variables $u$ and $z$ defined in Eq.\ (\ref{u.z.def}).  This
    gives us an overall factor, just like the $g^{2}/16\pi ^{2}$ in the
    self energy, times an integral over roughly the unit square
    in $u$ and $z$.

\end{itemize}

\begin{figure}
\cfig{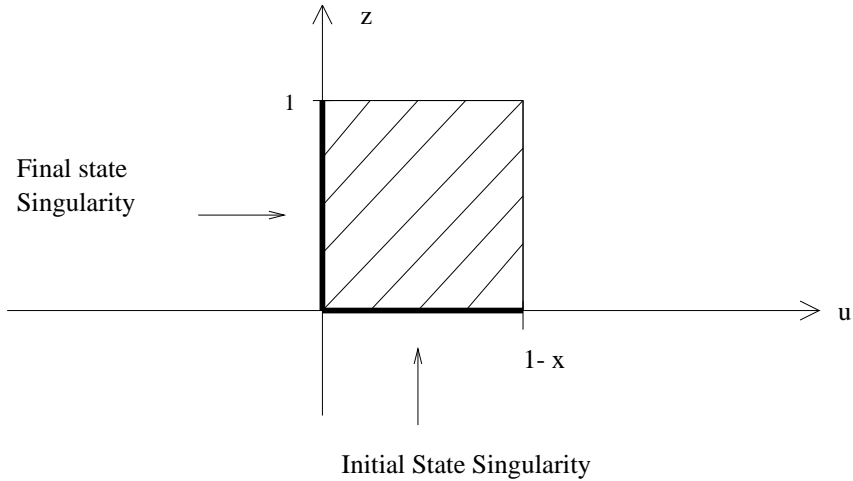}{2.5in}
\caption{Integration region and position of singularities for
    Fig.\ \protect\ref{fig:1-loop.DIS}(a).}
\label{fig:1-loop.Sing}
\end{figure}

\subsection{Real gluon}

For the real gluon graph Fig.\ \ref{fig:1-loop.DIS}(a) it is well
known that the integrand has the following singularities in the massless
limit:
\begin{itemize}

\item Initial-state collinear singularity on $z=0$.

\item Final-state collinear singularity on $u=0$.

\item Soft singularity $k=0$ at the intersection of the previous
    two singularities, i.e., at $u=z=0$.

\end{itemize}
(See Fig.\ \ref{fig:1-loop.Sing}.)  In accordance with the
standard recipe for constructing the Wilson coefficient, we
subtract the initial-state collinear singularity.
The subtraction term itself has a UV divergence, which
we choose to cancel by using the same method as
we used for the UV divergence of
the self-energy graph.
This gives\footnote{
  We have chosen the overall normalization of the graphs to be
  such that the lowest-order Born graph gives just $f(x)$.
}
\begin{equation}
   \frac {g^{2}}{8\pi ^{2}} C_{F} \int _{0}^{1} dz \int _{0}^{1-x} du
\,
f\left( \frac {x}{1-u} \right) \;
   \left[
        \frac {1-z}{zu}
        - \frac {\theta (z<z_{\rm cut})}{zu}
   \right] .
\label{Graph.a}
\end{equation}
The cut-off $z_{\rm cut}$ on the subtraction term is analogous to
the cut-off $\mu _{c}$ we used for the UV counterterm for the
self-energy graph.
The value of $z_{\rm cut}$ needed to reproduce the \MSbar{}
prescription can be found by a simple calculation from the
Feynman rules for parton densities.  But we will not use this
result here.  Rather, we will aim at calculating an appropriate
value for $z_{\rm cut}$ to keep the one-loop correction down to a
``normal size'' (and, most importantly, whether it is possible to
find such an appropriate value at all).
This effectively amounts to a choice of factorization scheme.
Once a suitable value for $z_{\rm cut}$
has been obtained, it is a mechanical matter to translate
it to a value for $\mu _{\MSbar}$ (or to a value of the scale $\mu $ in
any other chosen scheme).  The calculation may also result in a
need for an extra {\em finite} counterterm.

The subtraction in Eq.\ (\ref{Graph.a}) has evidently
accomplished its purpose of cancelling the initial-state
singularity.  But we are still left with the singularity on the
line $u=0$.
This singularity will cancel against a singularity in the virtual
graph, as we will now see.

\subsection{Sum of virtual and real graphs}

Next we compute the virtual graph of Fig.\ \ref{fig:1-loop.DIS}(b),
following the same line as in the previous subsection.
We will construct an integral in the same variables as the real
graph.  The reason why we do this is that the cancellation of
the divergence at $u=0$ will be point-by-point in the integrand.
This can be seen from the proof by Libby and Sterman \cite{LS}.
They treat a general case of final-state interactions, of which
our example is a particular case. They first treat one
integration analytically, with the aid of the mass-shell
conditions for the final state, in such a way that the
integrations for graphs related by different positions of the
final-state cut then have the same dimensions. After that, the
cancellation between the different graphs is point-by-point in
the integrand.

This implies that we need to perform the $\xi $ and ${\bf k}_{T}$ integrals.
We do the convolution with $\xi $ by the mass-shell $\delta $-function,
which now gives $\xi =x$.
Then we do the
${\bf k}_{T}$-integral analytically. (This is not the most trivial
integral, but it works conveniently with our choice of variables.)
An example of the type of integral encountered is:
\begin{equation}
 \int _{0}^{\infty }\frac {dk_{T}^{2}}{[Q^{2}(z-1)u-k_{T}^{2}+i\epsilon ] \,
[Q^{2}(u-1)z-k_{T}^{2}+i\epsilon ] \, [Q^{2}uz-k_{T}^{2}+i\epsilon ]} .
\nonumber
\end{equation}
In the virtual case, no longer does a positive energy condition restrict
the range of $u$ and $z$.  Nevertheless, it is convenient to split up the
result into a piece inside the square $0 \leq  u,z \leq  1$ and a piece
from outside the square.  It is sufficient to examine the
contribution within the square.  After subtracting the collinear
singularity and adding the result from the real graph we
get\footnote{
    In (\ref{Graph.a}) the upper limit on $u$ is $1-x$, but
    we replace the limit by $1$ in when we copy the formula into
    (\ref{Wilson.1}).  This
    change is innocuous since the limit $1-x$ arises from the fact
    that the parton density $f(\xi )$ is 0 when $\xi  > 1$, and
    hence that $f(x/(1-u))$ is 0 when $u > 1-x$. The limits on
    $u$ that result from positivity of energy of the two
    final-state lines in Fig.\ \ref{fig:1-loop.DIS}(a) are
    just $0 < u < 1$.
}
\begin{eqnarray}
   && \frac {g^{2}}{8\pi ^{2}} C_{F} \int _{0}^{1} \int _{0}^{1} du \, dz \;
   \left[ f\left( \frac {x}{1-u} \right) - (1-u)f(x)
   \right]
   \; \left[
        \frac {1-z}{zu} - \frac {\theta (z<z_{\rm cut})}{zu}
   \right]
\nonumber\\
    &&
    + \ \mbox{contribution from {\em outside} $0 \leq  u,z \leq  1$} .
\label{Wilson.1}
\end{eqnarray}
The UV divergence ($k\to \infty $) is from outside the square and is
cancelled by subtractions just as for the self-energy graph.

\begin{figure}
\cfig{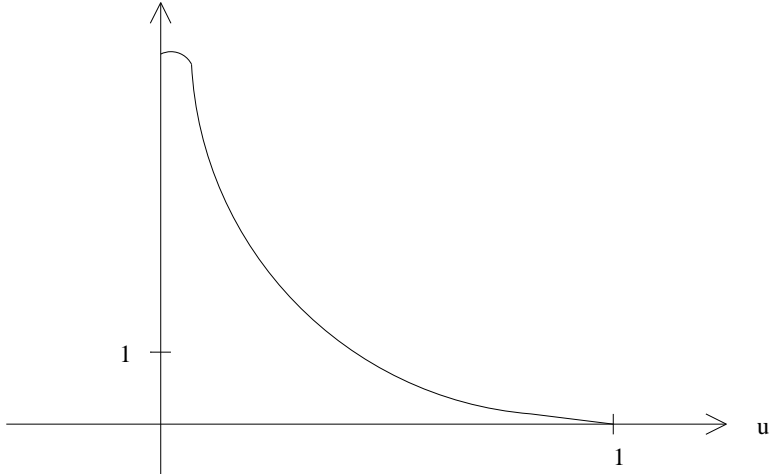}{2.5in}

\caption{The integrand of Eq.\ (\protect\ref{Wilson.1}) as a
         function of $u$ at fixed $z$ when $f(x)$ is a steeply
         falling function of $x$.
}
\label{fig:integrand}
\end{figure}

We see that the singularity at $u=0$ for fixed $z$ has cancelled,
even at the intersection of the two singular lines.
However, how good the cancellation is depends on the test
function.  If $f(x)$ is slowly varying, the cancellation is good
over a wide range of $u$.  But if $f(x)$ is steeply falling as
$x$ increases, the cancellation is good only over a narrow range
of $u$, and we are left with an integrand that behaves like $1/u$
for larger $u$.
For fixed $z$ we then have an integrand like
that in Fig.\ \ref{fig:integrand}.

This kind of behavior is quite typical when singularities are
cancelled between different graphs with different final states.
The integrand is a non-trivial distribution, and how good the
cancellation of the singularity is depends on properties of the
test function. The cancellation only occurs after integration
over a range of final states.  This is in distinct contrast with
the case of the initial-state singularity for which we have
constructed an explicit subtraction.  The cancellation of the
$z \to  0$ singularity occurs {\em before} the integration with the
test function.

Immediately we also get complications in computing the typical
size of the graph.
In particular, when $f(x)$ is steeply falling, we should
therefore expect a size for the integral that is much larger than
the natural size we defined earlier.

\subsection{Estimate of size}
\label{subsec:estimate}
First, as a benchmark case, let us assume that $f(x/(1-u))$ is slowly
and smoothly varying when $u$ increases from 0 to $1/2$, and
that\footnote{
   \label{z.equals.half}
   The reason for using `$u=1/2$' in these criteria rather than,
   say, $u=1$ is the same as for using $\mu =m$ in the calculation
   of the self-energy.  It is a rough attempt to optimize the
   errors without using the details of the integrand, since
   $u=1/2$ is midway between the singularity at $u=0$ and the
   approximate edge of the region of integration.  The
   integration region for the real graph extends to $u=1-x$,
   while that for the virtual graph extends beyond $u=1$.
   Analogous reasoning applies to $z$ and the relation between
   the real graph and its eikonal approximation even though, in this
   case, one has to look at the whole integrand since both test functions
   are of the form $f \left (\frac {x}{1-u} \right )$.
}
$z_{\rm cut}$ is around $1/2$.  Then the size of
(\ref{Wilson.1}) can be estimated as
\begin{equation}
   \frac {g^{2}}{8\pi ^{2}}
   \times  \left( \mbox{group theory} \right)
   \times  \left( \mbox{area of unit square in $(u,z)$} \right)
   \times  f(x)
   \simeq \frac {g^{2}}{6\pi ^{2}} f(x) .
\label{Wilson.nat}
\end{equation}
A way of obtaining this result with the same method as we used
for the self-energy is to write
\begin{equation}
   \frac {g^{2}}{(2\pi )^{4}} C_{F}
   \times  \Bigg( \mbox{range of $k$, i.e., $2\pi ^{2}Q^{4}$}
     \Bigg)
   \times  \Bigg( \mbox{size of integrand, $\displaystyle \frac {f(x)}{Q^{4}}$}
     \Bigg) .
\end{equation}
Given that the lowest-order graph is $f(x)$, and that the
integrand varies in sign, all this implies that the contribution
of this graph (with all the cuts and subtractions) is probably
somewhat smaller than $g^{2}/6\pi ^{2}$ times the lowest-order graph.
In other words, the simplest estimate for the graphs, Eq.\
(\ref{natural.size.QCD}), is a valid estimate in this situation.

There are a modest number of graphs, so this result would have
very nice implications for the good behavior of perturbation
theory: the real expansion parameter of QCD would be $\alpha _{s}/\pi $,
which is a few per cent in many practical situations.

Unfortunately, the conclusion is vitiated when $f(x)$ is steep,
as is often the case.  Consider the parton density factor times
the $1/u$ factor, relative to the lowest-order factor $f(x)$.  In
the limit $u \to  0$, this is
\begin{equation}
   \frac {f \left( \frac {x}{1-u} \right) - f(x)}{u f(x)}
   \sim \frac {xf'(x)}{f(x)} .
\label{u.to.0}
\end{equation}
This factor should be of order unity, if the previous estimate of
the size, Eq.\ (\ref{Wilson.nat}), is to be valid.

But if the logarithmic derivative $xf'(x) / f(x)$ is much bigger
than 1, then we have to change our estimates.  Consider a
typical ansatz for a parton density:
\begin{equation}
   f(x) \propto  (1-x)^{6} ,
\end{equation}
for which
\begin{equation}
   \left|
        \frac {xf'}{f}
   \right|
   = \frac {6x}{1-x} .
\end{equation}
This is 6 when $x=1/2$, and goes to infinity as $x \to 1$.
Clearly our estimate in Eq.\ (\ref{Wilson.nat}) is bad.  Moreover
the $u \to 0$ estimate, Eq.\ (\ref{u.to.0}), is only approximately
valid when $f(x)$ does not change by more than a factor 2
(roughly).  Once $u$ gets larger than the inverse of the
logarithmic derivative, the $f(x/(1-u))$ term
in Eq.\ (\ref{Wilson.1})
is no longer
important, and we get the result pictured in Fig.\
\ref{fig:integrand}: we have basically a
$1/u$ form with a cut-off at small $u$.  This is a recipe for a
large logarithm, with the argument of the logarithm being the
large logarithmic derivative.

To make the estimate, it is convenient to define
\begin{equation}
    \delta u = -\frac {f(x)}{2f'(x)x} .
\label{div.u}
\end{equation}
{}From Eq.\ (\ref{u.to.0}), we see that $\delta u$ is approximately the
change in $u$ to make $f(x/(1-u))$ fall by a factor of 2.
We interpret $\delta u$ as the value of $u$ at which the final-state
cancellations become ``bad''.

Next we note that for normal parton densities $f(x)$ is a
decreasing function of $x$.  So a simple useful estimate can be
made by making the following approximation:
\begin{equation}
    f\left ( \frac {x}{1-u} \right ) \simeq
    \left\{
    \begin{array}{l l}
        \displaystyle
        f(x)\left [1-\frac {u}{2\delta u}\right ] & {\rm if~} u < 2\delta u ,
    \\
        0                        & {\rm if~} u > 2\delta u
    \end{array}
    \right . .
\label{exp.u}
\end{equation}
Therefore, in Eq.\ (\ref{Wilson.1}), we
can replace $\int _{0}^{1} \frac {du}{u}f(x/(1-u))$ by
$\int _{0}^{2\delta u} \frac {du}{u} f(x)$.

Our estimate for Eq.\ (\ref{Wilson.1}) is the sum of
contributions from the following regions:
\begin{itemize}

\item
    $0<u<2\delta u$:
    The value of the integrand is about $-f(x)/2\delta u$ times a
    function of $z$.  For good choices of $z_{\rm cut}$, the
    function of $z$ is less than about unity, but with an
    indefinite sign.  Thus we obtain a contribution of about
    $f(x)$ in size.

\item
    $2\delta u<u<1$:
    The integrand is now approximately $-f(x)/u$,
    again times a mild function of
    $z$.  We therefore obtain a contribution of about
    $f(x)\ln(1/2\delta u)$.

\item
    Exterior of unit square:
    Here, the only contribution is from the virtual graph, and we
    have no final-state singularity. Hence the naive estimate of
    unity is valid.
    (The precise value, when $\mu =Q$, is in fact somewhat larger.)

\end{itemize}
All of these are to be multiplied by the pre-factor $g^{2}C_{F}/8\pi ^{2}$.
So we obtain a total contribution of
\begin{equation}
   \frac {g^{2}}{8\pi ^{2}} C_{F} f(x) \left[ \pm 2 \pm  \ln \left( \frac
{1}{2\delta u} \right)
               \right] ,
\label{estimate.vertex}
\end{equation}
where each term represents an estimate, valid up to a factor of 2
or so.  The contributions from within the unit square may have either
sign, depending on the cut in the collinear subtraction, while
the contributions from outside the unit square, from virtual
graphs only, have a negative sign.
As an explicit indication that our estimates are valid for the
sizes but not the signs of the graphs, we have inserted a $\pm $
sign in front of each term.
This estimate assumes that renormalization of the UV divergence
of the virtual graph is done at the natural scale $\mu  \simeq Q$,
and that renormalization of the parton densities is done so that
it corresponds to $z_{\rm cut}$ of about 1/2.
It also assumes that $f(x)$ falls steeply enough for $\delta u$ to be
less than about 1, as is typically true.

\subsection{Interpretation}

It is obvious that there is a logarithmic enhancement in Eq.\
(\ref{estimate.vertex}) whenever $\delta u$ is small. Our calculation
is, of course, no more than a rederivation of the standard
observation that there are large logarithms in the $x \to 1$ region.
What our derivation adds is to show that it is not so much the
limit $x \to 1$ that is causing the problem as the steepness of the
parton densities. Moreover we have given a numerical criterion
for when the correction begins to be larger than what we called
the natural size for higher-order corrections.  In other words we
have shown how to estimate the constant term that accompanies the
logarithm.  Moreover this is all presented in the context of a
general method for obtaining estimates of the sizes of graphs.

It is perhaps clear that, with sufficient foresight, one could
have predicted the large corrections merely from the observation
that the derivation of the factorization theorem requires the
cancellation of final-state divergences between different final
states.

There is in fact another source of large corrections that will
make its effect felt in even higher order.  This is a mismatch
in the scales needed for renormalizing the parton densities.  We
have renormalized these by using a value of
$z_{\rm cut}$ that must be about 1/2 to avoid making the
contribution of the graph unnecessarily large.  In the case of
the real graph, we can translate this to a scale of transverse
momentum by using the mass-shell condition for the gluon:
\begin{equation}
   k_{T}^{2} = Q^{2} u z (1-z) \frac {\xi }{x} = Q^{2} \frac {u z (1-z)}{1-u} .
\end{equation}
Evidently, whenever small values of $u$ are important, small
values of $k_{T}$ (relative to $Q$) will be important. This does not
affect our one-loop calculation. But in higher-order correction,
the virtuality of some internal lines will be controlled by the
value of $k_{T}$, and hence there will be mismatches between the
scales needed at different steps in the calculation.

Once one has diagnosed the problem, we see that a proper solution
lies in more accurately calculating the form of the Wilson
coefficient near its singularity.  This subject goes under the
heading of resummation of large corrections \cite{resum}.

\section{More detailed estimation of $F_{T,L}$ to one-loop order}
\label{sec:full.estimate}

In the following we will estimate the sizes of the
one-loop corrections to
the structure functions $F_{T}$ and $F_{L}$ {\em without} doing actual
calculations of Feynman diagrams by giving a recipe of how to construct
the estimates from general principles and kinematic considerations.
However, we will present the recipe in the context of an actual set of
Feynman graphs.
By using the calculations of the graphs in the appendix, we will
verify that these ``simple-minded'' estimates are actually valid.

One obtains $F_{T,L}$ from the well-known hadronic tensor
$W^{\mu \nu }$ by
projecting out the ``transverse'' and ``longitudinal'' pieces via:
\begin{eqnarray}
F_{T} = -g_{\mu \nu }W^{\mu \nu }&=& 3F_{1} - \frac {F_{2}}{2x} \nonumber\\
F_{L} = \frac {Q^{2} p_{\mu }p_{\nu }}{p\cdot q^{2}}W^{\mu \nu }&=& -F_{1} +
\frac {F_{2}}{2x} ,
\end{eqnarray}
which give, for example,
\begin{equation}
F_{1} = \frac {1}{2} \left [-g_{\mu \nu } - \frac {Q^{2} p_{\mu }p_{\nu
}}{p\cdot q^{2}} \right ]W^{\mu \nu } .
\end{equation}

\subsection{Estimation of $F_{L}$}

The recipe for estimating $F_{L}$ is the following:
\begin{itemize}

\item The singularities we encounter (UV, collinear and soft) are
      all in the
      form of a factor times the Born graph, and the Born graph has no
      longitudinal part. Therefore the one-loop graphs for $F_{L}$ have no
      UV, soft or collinear singularities.

\item Since the parton densities are falling with increasing $x$, the size
      of a graph is:
      \begin{equation}
        \frac {g^{2}}{8\pi ^{2}}\times C_{F}\times f(x)\times \mbox{{\rm range
of $u$}}
        \times \mbox{{\rm range of $z$}}.
      \end{equation}

\item The range of $z$ is 1.

\item The range of $u$ is $\delta u$.

\end{itemize}
It is elementary to show that the self-energy and vertex graphs
(whether real or virtual) give a zero contribution to $F_{L}$.
Therefore, our result for $F_{L}$ is:
\begin{equation}
    F_{L} = \frac {g^{2}}{6\pi ^{2}}f(x)\delta u .
\label{FL.estimate}
\end{equation}

Let us now check whether our intuition has guided us in the right way.
We use Eq.\ (\ref{FL.ladder}) for the contribution to the
coefficient function for $F_{L}$.  We approximate the $u$ integral
by
\begin{equation}
   \int  du f\left( \frac {x}{1-u} \right)
   \simeq \delta u f(x) ,
\end{equation}
which is appropriate for a typical parton density, which falls
with increasing $x$.  The $z$ integral gives a factor $1/2$, and
we recover Eq.\ (\ref{FL.estimate}), which we obtained by more
general arguments.

Notice that $F_{L}$ is generally rather smaller than what we have
termed the natural size, because of the $\delta u$ factor: i.e.,
because of the restricted phase space available.  There are no
enhancements due to final-state singularities.

Our estimation methods can be applied in two ways.  One is to
estimate the graphs without having explicit expressions for the
graphs; one just searches for the possible singularities that
would prevent the natural size of a graph from being its actual
size.  The second way of using the methods is to examine the
expressions for graphs with the knowledge of the singularity and
subtraction structure and to perform a more direct estimate of
the sizes.  This is useful since the integrals must often be
performed numerically, e.g., whenever one convolutes with parton
densities that are only known numerically. Additional information
that is now obtained concerns the typical virtualities, etc., of the
internal lines of the graphs (compare Neubert's work
\cite{Neubert}).  This enables a diagnosis to be made of the
extent to which a problem is a multi-scale problem and therefore
in need of resummation.

A point that we have not addressed is the estimation of the size
of the trace of a string of gamma matrices, for example.  In the
standard formula for such a trace involves a large number of
terms.  Nevertheless, it is evident from the above calculations
that there are cancellations.  The final result is that we
obtain, relative to corresponding numbers for a scalar field
theory, a small factor (1 to 4) times a standard
Lorentz-invariant quantity for the process.

Evidently, more work on this subject is needed.  But the issue of
the size of the numerator factors from traces, etc., affects
completely finite quantities, such as the one-loop coefficient
function for $F_{L}$, just as much as quantities with divergences
that are cancelled.  However, it is the latter quantities that
have the potential for especially large corrections.  Numerator
factors result in magnitudes common to all graphs.

\subsection{Estimation of $F_{T}$}

We have already calculated one graph, Fig.\ \ref{fig:1-loop.DIS},
for $F_{T}$, so we only need to summarize the method and apply it to
the remaining graphs.  The general procedure is:
\begin{itemize}

\item Since we have singularities in our graphs, they have to be
    cancelled.  In Wilson coefficients such as we are
    calculating, there are explicit subtractions for the collinear
    initial-state singularities.  Then there are explicit
    subtractions for UV divergences (both those associated with
    the interactions and those needed to define the parton
    densities and that therefore enter into the initial-state
    subtractions).  Finally there are final-state singularities
    that cancel between real and virtual graphs.

\item We consider separately the integrations inside and outside
    the square $0 < u, z <1$.

\item A term of the order of the natural size arises from outside the
    square $0 < u, z < 1$.  This comes from purely virtual
    graphs, with their collinear subtraction.

\item Inside the square, a real graph without a final-state
    singularity contributes an amount of the order of the natural size
    times $\delta u$.  This reflects the restriction on the range of
    integration imposed by the parton density.

\item Similarly a virtual graph without a final-state singularity
    contributes a term of the natural size.

\item Finally, the sum of real and virtual graphs with a
    final-state singularity contributes a term of the order of the
    natural size enhanced by a factor $2 + \ln (1/2\delta u)$, just as
    in our estimate of Fig.\ \ref{fig:1-loop.DIS}.

\end{itemize}

We have the graphs of Fig.\ \ref{fig:1-loop.DIS}, their Hermitian
conjugates, the cut and uncut self-energy graph of Fig.\
\ref{fig:qself.energy}, below, and the ladder graph, Fig.\
\ref{fig:remaining.graph}.  Since $\delta u \lesssim 1 $ typically, the
ladder graph gives a small contribution, and it is sufficient to
multiply the estimate of the Fig.\ \ref{fig:1-loop.DIS} by 3:
\begin{equation}
F_{T} = \frac {g^{2}}{2\pi ^{2}} f(x) \left[ 2 + \ln \left( \frac {1}{2\delta
u} \right)
               \right] .
\label{eq.est}
\end{equation}

To see how this compares with the results from the actual graphs, we
use those in the appendix.

The cut and uncut self-energy graphs, Fig.\
\ref{fig:qself.energy} below, have final-state singularities, as
can be seen in Eq.\ (\ref{FT.se}).
We found it convenient to use $z$ and $k_{T}$ as integration
variables.  There is a singularity at $k_{T}=0$ in each individual
graph.  In analogy to the definition of $\delta u$, Eq.\ (\ref{div.u}),
we define
\begin{equation}
   \delta k_{T}^{2} = \frac {-Q^{2} f(x)}{2 f'(x)},
\end{equation}
and by following the same steps as we applied to the ladder
graph, we find the following estimate for the contribution of
Fig.\ \ref{fig:qself.energy}:
\begin{equation}
   - \frac {g^{2}}{16\pi ^{2}} C_{F} f(x) \left[ 1 + \ln \left( \frac {\mu
^{2}}{2\delta k_{T}^{2}} \right)
               \right] .
\label{se.estimate}
\end{equation}
We have inserted a $-$ sign in this estimate; it is fairly easy
to see that the coefficient of the logarithm is negative.

For the ladder graph, whose contribution to $F_{T}$ is in Eq.\
(\ref{FT.ladder}), there is only an initial-state singularity and
that is cancelled by an explicit subtraction.  If $z_{\rm cut}$ is
around $1/2$, then we can apply the same reasoning as for the
longitudinal part of the ladder graph, and we find an estimate
\begin{equation}
    \pm \frac {g^{2}}{6\pi ^{2}}f(x)\delta u ,
\label{ladder.T.estimate}
\end{equation}
where, as in Eq.\ (\ref{estimate.vertex}), we use the $\pm $ to
indicate that our estimates do not determine the sign of the
contribution.

We have already examined the cut and uncut vertex graphs, in
Eq.\ (\ref{estimate.vertex}).  When we multiply this by 2 (to
allow for the Hermitian conjugate graphs) and add the self-energy
and ladder contributions, Eqs.\ (\ref{se.estimate}) and
(\ref{ladder.T.estimate}), we get somewhat less than our original
estimate, Eq.\ (\ref{eq.est}), provided the renormalization
mass $\mu $ is in a reasonable range.  This lower value is
because the self-energy graph is simpler than the vertex graph.

\section{Conclusions}
\label{sec.conclusions}

\begin{itemize}

\item We have a systematic method for estimating the sizes of
    higher-order graphs.

\item The natural expansion parameter in QCD is of the order of
    \begin{equation}
       \frac {g^{2}}{8\pi ^{2}} \times  \mbox{group theory}
       \times  \mbox{number of graphs}.
    \end{equation}
    In answer to a question asked when this work was presented at
    a conference, let us observe: {\em The above number may not
    be the actual expansion parameter, but we argue that this
    natural size sets a measure of whether actual higher-order
    corrections are of a normal size or are especially large. }

\item The method allows an identification of appropriate sizes
    for renormalization and factorization scales.

\item By an examination of the kinematics of graphs,
    we can identify contributions that are large compared with
    the estimates based merely on the size of the natural
    expansion parameter.

\item Our method should give a systematic technique of diagnosing
    the reasons for the large corrections, and hence of
    indicating where one should work out resummation methods.

\item The method gives an algorithm for the numerical integration
    of graphs, both real and virtual.

\end{itemize}

This is clearly just the start of a project, and there is also
a lot of overlap with work such as that of Catani and Seymour
\cite{Catani.Seymour}, of Brodsky and Lu \cite{BL}, and of
Neubert \cite{Neubert}.

\section*{Acknowledgements}

This work was supported in part by the U.S. Department of Energy
under grant number DE-FG02-90ER-40577.


\appendix

\section{The remaining longitudinal and transverse contributions
         to the Wilson coefficient}
\label{sec:appendix}

\subsection{Transverse part of the self-energy}

\begin{figure}
\cfig{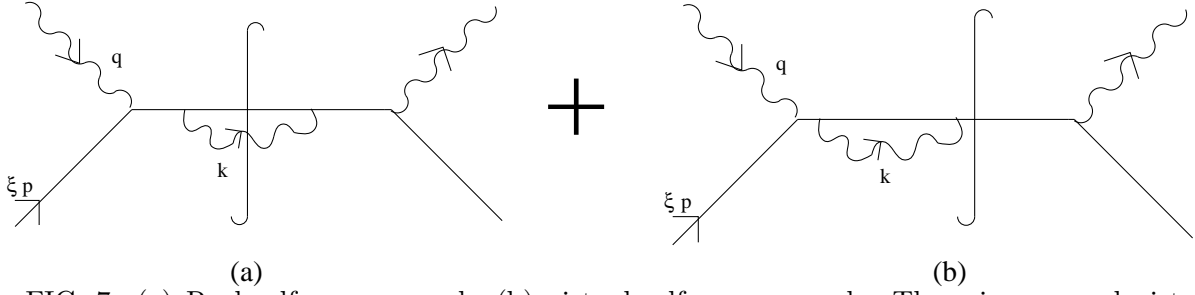}{1.5in}

\caption{(a) Real self-energy graph, (b) virtual self-energy graph.
    There is a second virtual graph that is the conjugate of
    graph (b).
}
\label{fig:qself.energy}
\end{figure}

The real part of the self-energy is computed along the same lines
as mentioned in the main body of the text, except that we chose to
integrate over $u$ instead of $k_{T}$ as a matter of convenience.
The graph and its possible cuts are shown in Fig.\
\ref{fig:qself.energy}.
As far as the virtual part is concerned, one notes first that the general
structure of the quark self-energy for zero mass is \cite{Pokorski}:
\begin{equation}
  \Sigma (\hat {p}) = B(0)\hat {p} ,
\end{equation}
where $B(0)$ is computed via:
\begin{equation}
               B(0) = \frac {1}{4}\left. {\rm Tr}\left(
                  \gamma ^{+}\frac {\partial  \Sigma(\hat p)}{\partial p^{+}})
                \right )   \right \vert _{\hat p=0} .
\end{equation}
The result for the virtual graph is then given by simply multiplying the
Born result by $B(0)$ and convoluting with the test function.
The complete result according to our prescription is:
\begin{equation}
\frac {g^{2}}{8\pi ^{2}} C_{F}\left[ \int ^{1}_{0} dz \int ^{k^{2}_{T,max}}_{0}
d{\bf k}_{T}^{2} \frac {z}{k_{T}^{2}}f\left (\frac {x}{y}\right )
- \int ^{1}_{0} dz \int ^{\mu ^{2}}_{0} d{\bf k}_{T}^{2} \frac {z}{k_{T}^{2}}
f(x) \right ] ,
\label{FT.se}
\end{equation}
where $\frac {1}{y} = 1 + \frac {k_{T}^{2}}{Q^{2}}$, after having made
the change of variable $k_{T}^{2} \rightarrow k_{T}^{2}z(1-z)$,
which also gives $k^{2}_{T,max} = \frac {Q^{2}(1-x)}{x}$.
Renormalization by subtraction of the asymptote is used in our
formula with a cut-off on the momentum integral for the virtual
graph of $\mu ^{2}$.
As one can see there are no initial-state singularities simply because
propagator corrections do not induce initial-state singularities as vertex
corrections do. The final-state singularities are still there;
thus one needs
to look at both the real and the virtual graphs together.

\subsection{The transverse part of the ladder diagram}

\begin{figure}
\cfig{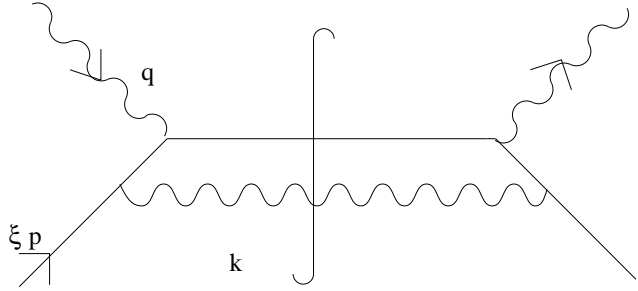}{1.5in}

\caption{The ladder graph for the Wilson coefficient.}
\label{fig:remaining.graph}
\end{figure}

The calculation of the ladder graph has been carried out as
outlined in the main text, yielding:
\begin{eqnarray}
\frac {g^{2}}{8\pi ^{2}} C_{F}\int ^{1-x}_{0} du \int ^{1}_{0} dz
\frac {u}{ z(1-u)}
f\left (\frac {x}{1-u}\right ) \left [1 - \theta (z - z_{\rm cut})
\right ] .
\label{FT.ladder}
\end{eqnarray}
The second term, with its $\theta $-function, is the collinear
subtraction, whose UV divergence is cancelled by subtraction of
the asymptote.  The above formula assumes that $z_{\rm cut} < 1$.
If $z_{\rm cut} > 1$, then we must extend the $z$ integration in the
second term beyond the limit $z=1$, of course.

\subsection{Longitudinal part of the vertex correction, self-energy and
ladder diagram}

The calculation has been carried out as outlined in the main text and
yields the following for the ladder graph:
\begin{equation}
\frac {g^{2}}{4\pi ^{2}} C_{F}\int ^{1-x}_{0} du \int ^{1}_{0} dz
(1-z) f\left (\frac {x}{1-u}\right ) .
\label{FL.ladder}
\end{equation}
The real and virtual graphs give $0$ for both the self-energy and the vertex
correction. There is no virtual graph for the ladder diagram.


\end{document}